\documentclass[12pt,english,reprint,aps]{revtex4}
\usepackage[T1]{fontenc}
\usepackage[latin9]{inputenc}
\usepackage[a4paper]{geometry}
\usepackage{float}
\usepackage{mathtools}
\usepackage{amsmath}
\usepackage{amssymb}
\usepackage{graphicx}

\makeatletter
\@ifundefined{textcolor}{}
{%
 \definecolor{BLACK}{gray}{0}
 \definecolor{WHITE}{gray}{1}
 \definecolor{RED}{rgb}{1,0,0}
 \definecolor{GREEN}{rgb}{0,1,0}
 \definecolor{BLUE}{rgb}{0,0,1}
 \definecolor{CYAN}{cmyk}{1,0,0,0}
 \definecolor{MAGENTA}{cmyk}{0,1,0,0}
 \definecolor{YELLOW}{cmyk}{0,0,1,0}
}

\usepackage{lipsum}

\usepackage{babel}

\makeatother

\usepackage{babel}
\begin{document}

\title{Structural scale $q-$derivative and the LLG-Equation in a scenario
with fractionality}

\author{J.Weberszpil}

\email{josewebe@gmail.com}

\affiliation{Universidade Federal Rural do Rio de Janeiro, UFRRJ-IM/DTL}

\affiliation{Av. Governador Roberto Silveira s/n- Nova Igua\c{c}\'u, Rio de Janeiro,
Brasil, 695014.}

\author{J. A. Helay\"el-Neto}

\email{helayel@cbpf.br}

\address{Centro Brasileiro de Pesquisas F\'isicas-CBPF-Rua Dr Xavier Sigaud
150,}

\address{22290-180, Rio de Janeiro RJ Brasil.}
\begin{abstract}
{\normalsize{}In the present contribution, we study the Landau-Lifshitz-Gilbert
equation with two versions of structural derivatives recently proposed:
the $scale-q-$derivative in the non-extensive statistical mechanics
and the axiomatic metric derivative, which presents Mittag-Leffler
functions as eigenfunctions. The use of structural derivatives aims
to take into account long-range forces, possible non-manifest or hidden
interactions and the dimensionality of space. Having this purpose
in mind, we build up an evolution operator and a deformed version
of the LLG equation. Damping in the oscillations naturally show up
without an explicit Gilbert damping term.}{\normalsize \par}
\end{abstract}

\date{\today}

\maketitle
\medskip{}


\section{Introduction}

In recent works, we have developed connections and a variational formalism
to treat deformed or metric derivatives, considering the relevant
space-time/ phase space as fractal or multifractal \cite{Nosso On conection2015}
and presented a variational approach to dissipative systems, contemplating
also cases of a time-dependent mass \cite{Variational Deformed}.

The use of deformed-operators was justified based on our proposition
that there exists an intimate relationship between dissipation, coarse-grained
media and a limit energy scale for the interactions. Concepts and
connections like open systems, quasi-particles, energy scale and the
change in the geometry of space\textendash time at its topological
level, nonconservative systems, noninteger dimensions of space\textendash time
connected to a coarse-grained medium, have been discussed. With this
perspective, we argued that deformed or, we should say, Metric or
Structural Derivatives, similarly to the Fractional Calculus (FC),
could allows us to describe and emulate certain dynamics without explicit
many-body, dissipation or geometrical terms in the dynamical governing
equations. Also, we emphasized that the paradigm we adopt was different
from the standard approach in the generalized statistical mechanics
context \cite{Tsallis1,Tsallis BJP- 20 anos,Tsallis2}, where the
modification of entropy definition leads to the modification of the
algebra and, consequently, the concept of a derivative \cite{Nosso On conection2015,Variational Deformed}.
This was set up by mapping into a continuous fractal space \cite{Balankin PRE85-Map-2012,Balankin Rapid Comm,Balankin-Towards a physics on fractals}
which naturally yields the need of modifications in the derivatives,
that we named deformed or, better, metric derivatives \cite{Nosso On conection2015,Variational Deformed}.
The modifications of the derivatives, accordingly with the metric,
brings to a change in the algebra involved, which, in turn, may lead
to a generalized statistical mechanics with some adequate definition
of entropy.

The Landau-Lifshitz-Gilbert (LLG) equation sets out as a fundamental
approach to describe physics in the field of Applied Magnetism. It
exhibits a wide spectrum of effects stemming from its non-linear structure,
and its mathematical and physical consequences open up a rich field
of study. We pursue the investigation of the LLG equation in a scenario
where complexity may play a role. The connection between LLG and fractionality,
represented by an $\alpha-$deformation parameter in the deformed
differential equations, has not been exploited with due attention.
Here, the use of metric derivatives aims to take into account long-range
forces, possible non-manifest or hidden interactions and/or the dimensionality
of space.

In this contribution, considering intrinsically the presence of complexity
and possible dissipative effects, and aiming to tackle these issues,
we apply our approach to study the LLG equation with two metric or
structural derivatives, the recently proposed scale $-q-$derivative
\cite{Variational Deformed} in the nonextensive statistical mechanics
and, as an alternative, the axiomatic metric derivative (AMD) that
has the Mittag-Leffler function as eigenfunction and where deformed
Leibniz and chain rule hold - similarly to the standard calculus -
but in the regime of low-level of fractionality. The deformed operators
here are local. We actually focus our attention to understand whether
the damping in the LLG equation can be connected to some entropic
index, the fractionality or even dimensionality of space; in a further
step, we go over into anisotropic Heisenberg spin systems in (1+1)
dimensions with the purpose of modeling the weak anisotropy effects
by means of some representative parameter, that depends on the dimension
of space or the strength of the interactions with the medium. Some
considerations about an apparent paradox in the magnetization or angular
damping is given.

Our paper is outlined as follows: In Section 2, we briefly present
the scale$-q-$derivative in a nonextensive context, building up the
$q-$ deformed Heisenberg equation and applying to tackle the problem
of the LLG equation; in Section 3, we apply the axiomatic derivative
to build up the $\alpha-$deformed Heisenberg equation and to tackle
again the problem of LLG equation. We finally present our Conclusions
and Outlook in Section 4.

\section{Applying scale$-q-$derivative in a nonextensive context}

Here, in this Section, we provide some brief information to recall the main forms of scale $-q-$ derivative. The readers may see ref. \cite{Nosso On conection2015,Variational Deformed,Balankin PRE85-Map-2012} for more details.

Some initial claims here coincide with our work of Refs. \cite{Nosso On conection2015,Variational Deformed}
and the approaches here are in fact based on local operators \cite{Nosso On conection2015}.

The local differential equation,

\begin{equation}
\frac{dy}{dx}=y^{q},
\end{equation} with convenient initial condition, yields the solution given by the q-exponential, $y=e_{q}(x)$ \cite{Tsallis1,Tsallis BJP- 20 anos,Tsallis2}.

The key of our work here is the Scale$-q-$derivative (Sq-D) that we have recently defined as

\begin{equation}
{\displaystyle D_{(q)}^{\lambda}f(\lambda x)\equiv}{\displaystyle [1+(1-q)\lambda x]\frac{df(x)}{dx}.}\label{eq:q-derivative redef}
\end{equation}

The eigenvalue equation holds for this derivative operator, as the
reader can verify: 

\begin{equation}
D_{(q)}^{\lambda}f(\lambda x)=\lambda f(\lambda x).
\label{eq:Eigenvalue}
\end{equation}

\subsection{{\normalsize{}$q-$ deformed Heisenberg Equation in the Nonextensive
Statistics Context}}

With the aim to obtain a scale$-q-$ deformed Heisenberg equation, we now consider
the $scale-q-$ derivative \cite{Variational Deformed}

\begin{equation}
\frac{d^{q}}{dt^{q}}=(1+(1-q)\lambda x\frac{d}{dx}
\end{equation}
and the Scale -$q-$ Deformed Schr\"odinger Equation \cite{Variational Deformed},

\begin{equation}
i\hbar D_{q,t}^{\lambda}\psi=-\frac{\hbar^{2}}{2m}\nabla^{2}\psi-V\psi=H\psi,\label{eq:Schrodinger Scale-q}
\end{equation}
that, as we have shown in \cite{Variational Deformed}, is related
to the nonlinear Schr\"odinger equation referred to in Refs. \cite{NRT-PRL-2011}
as NRT-like Schr\"odinger equation (with $q=q'-2$ compared to the $q-$index
of the reference) and can be thought as resulting from a $time-scale-q-$deformed-derivative
applied to the wave function $\psi$.

Considering in eq.(\ref{eq:Schrodinger Scale-q}), $\psi(\vec{r,t})=U_{q}(t,t_{0})\psi(\vec{r,t}_{0})$,
the $q-$ evolution operator naturally emerges if we take into account
a $time-scale-q-$ deformed-derivative (do not confuse with formalism
of discrete scale time derivative): 

\begin{equation}
U_{q}(t,t_{0})=e_{q}^{(-\frac{i}{\hbar M_{q}}\mathcal{H}_{q}t)}.
\end{equation}

Here, $M_{q}$ is a constant for dimensional regularization reasons.
Note that the q-deformed evolution operator is neither Hermitian nor
unitary, the possibility of a $q-$unitary as $U_{q}^{\dagger}(t,t_{0})\otimes_{q}U_{q}(t,t_{0})=\mathbf{1}$ could be thought to come over these facts. In this work, we assume
the case where the commutativity of $U_{q}$ and $\mathcal{H}$ holds, but the $q-$unitarity is also a possibility.

Now, we follow similar reasonings that can be found in Ref.\cite{Zitter}
and considering the Sq-D.

So, with these considerations, we can now write a nonlinear $Scale-q-$deformed
Heisenberg Equation as 

\begin{equation}
D_{t,q}^{\lambda}\hat{A}(t)=-\frac{i}{\hbar M_{q}}[\hat{A},\mathcal{H}],\label{eq:q-heisenberg}
\end{equation} where we supposed that $U_{q}$ and $\mathcal{H}$ commute and $M_{q}$
is some factor only for dimensional equilibrium.

\subsection{{\normalsize{}$q-$deformed LLG Equation}}

To build up the scale$-q-$ deformed Landau-Lifshitz-Gilbert Equation, we consider eq.(\ref{eq:q-heisenberg}),  with $\hat{A}(t)=\hat{S}_{q}$

\begin{equation}
D_{t,q}^{\lambda}\hat{S}_{q}(t)=-\frac{i}{\hbar M_{q}}[\hat{S}_{q},\mathcal{H}],
\end{equation} where we supposed that $U_{q}$ and $\mathcal{H}$ commute.

\begin{equation}
\mathcal{H}=-g_{q}\frac{\mu_{B}}{\hbar M_{q}}\hat{S}_{q}\circ\vec{H}_{eff}.
\end{equation}

Here, $\vec{H}_{eff}$ is some effective Hamiltonian whose form that
we shall clearly write down in the sequel.

The scale$-q-$deformed momentum operator is here defined as $\widehat{p}_{q'}^{\lambda}=-i\hbar M_{q'}[1+\lambda(1-q')x]\frac{\partial^{q}}{\partial x^{q}}.$

Considering this operator, we obtain a deformed algebra, here in terms of commutation relation between coordinate and momentum

\begin{equation}
\left[\hat{x}_{i}^{q},\hat{p}_{j}^{q}\right]=\imath[1+\lambda(1-q')x]\hbar M_{q'}\delta_{\imath j}I
\end{equation}
and, for angular momentum components, as 
\begin{equation}
\left[\hat{L}_{i}^{q},\hat{L}_{j}^{q}\right]=\imath[1+\lambda(1-q')x]\hbar M_{q}\hat{L}_{k}^{q}.
\end{equation}

The $q'$ factor in $\hat{x}_{\imath}^{q'},\hat{p}_{j}^{q'},\hat{L}_{i}^{q'},\hat{L}_{j}^{q'},M_{q'}$
is only an index and $q$ is not necessarily equal to $q'$.

The resulting scale$-q-$deformed LLG equation can now be written as

\begin{equation}
D_{t,q}^{\lambda}\hat{S}_{q}(t)=-\frac{[1+\lambda(1-q')x]g_{q}\mu_{B}}{\hbar M_{q}}\hat{S_{q}}\times\vec{H}_{eff}.\label{eq:Scale-q-deformed LLG}
\end{equation}

Take $\hat{m}_{q}\equiv\gamma_{q}\hat{S}_{q},$ $\gamma_{q'}\equiv\frac{[1+\lambda(1-q')x]g_{q}\mu_{B}}{\hbar M_{q}}$.

If we consider that the spin algebra is nor affected by any emergent
effects, we can take $q'=1$.

Considering the eq.(\ref{eq:q-heisenberg}) with $\hat{A}(t)=\hat{S_{q}}$
and $\hat{m_{q}}=\left|\gamma_{q}\right|\hat{S_{q}}$ and $q'=1$;
we obtain the $q-$time deformed LLG dynamical equation for magnetization
as

\begin{equation}
D_{t,q}^{\lambda}\hat{m_{q}}(t)=-\left|\gamma\right|\hat{m_{q}}\times\vec{H}_{eff}.\label{eq:q-LLG equation for m}
\end{equation}

Considering $\vec{H}_{eff}=H_{0}\hat{k},$ we have the solution:

\begin{equation}
m_{x,q}=\rho\cos_{q}(\theta_{0})\cos_{q}(\gamma H_{0}t)+\rho\sin_{q}(\theta_{0})\sin_{q}(\gamma H_{0}t).
\end{equation}

In the figure, $\theta_{0}=0.$

\begin{figure}[H]
\includegraphics[width=8cm,height=8cm,keepaspectratio]{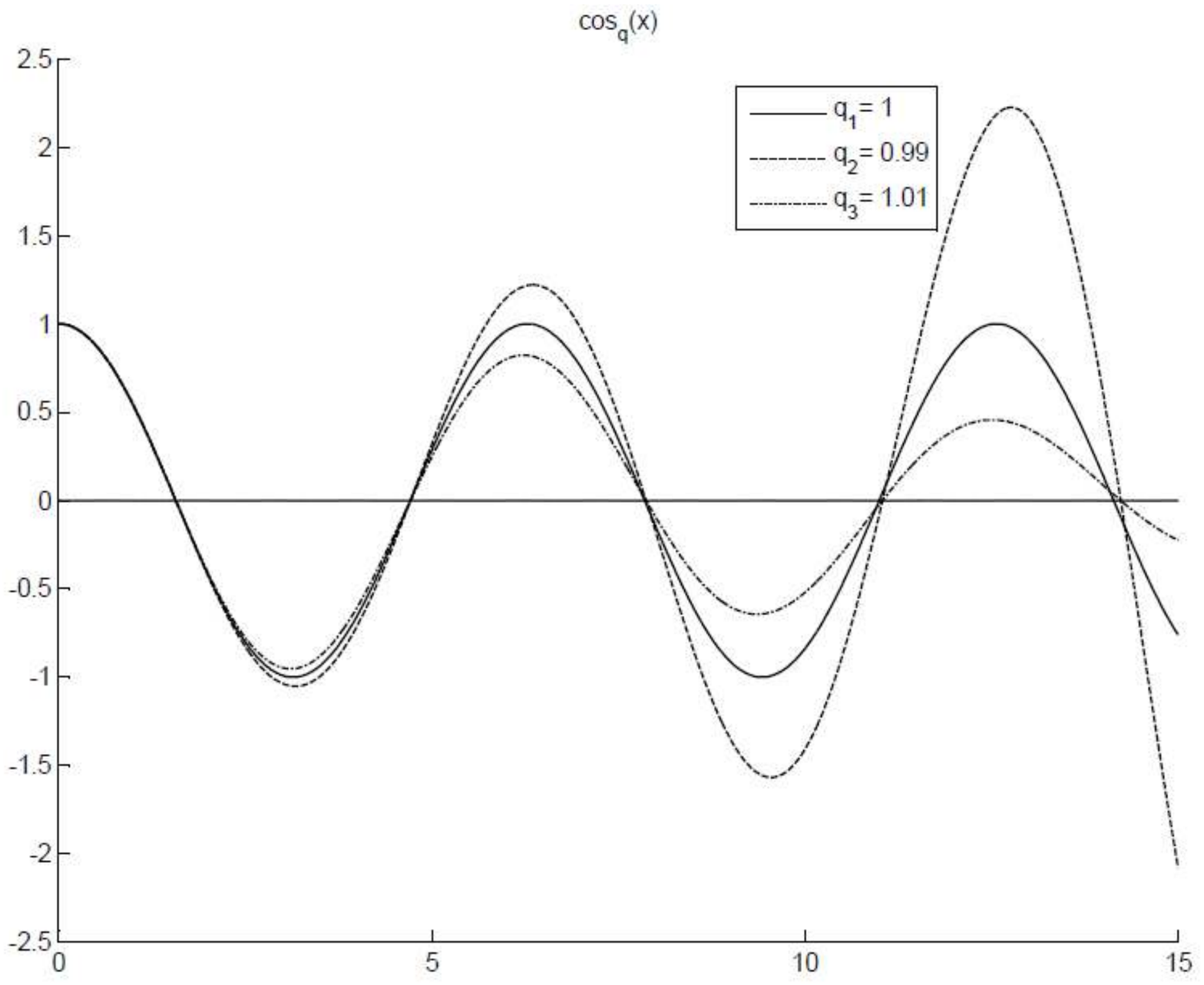}
\caption{Increase/Damping- $cos_{q}(x)$}. 
\end{figure}

\section{Applying Axiomatic Derivative and the $\alpha-$Deformed Heisenberg
Equation}
Now, to compare results with two different local operators, we apply
the axiomatic metric derivative.
Following the steps on \cite{Zitter} and considering the axiomatic
MD \cite{Axiomatic Metric}, there holds the eigenvalue equation $D_{x}^{\alpha}E_{\alpha}(\lambda x^{\alpha})=\lambda E_{\alpha}(\lambda x^{\alpha}),$
where $E_{\alpha}(\lambda x^{\alpha})$ is the Mittag-Leffler function
that is of crucial importance to describe the dynamics of complex
systems. It involves a generalization of the exponential function
and several trigonometric and hyperbolic functions. The eigenvalue
equation above is only valid if we consider $\alpha$ very close to
$1.$ This is what we call low-level fractionality \cite{Axiomatic Metric}.
Our proposal is to allow the use o Leibniz rule, even if it would
result in an approximation.
So, we can build up an\textbf{ }evolution operator:

\begin{equation}
U_{\alpha}(t,t_{0})=E_{\alpha}(-\frac{i}{\hbar^{\alpha}}\mathcal{H}t^{\alpha}),
\end{equation}
and for the deformed Heisenberg Equation

\begin{equation}
D_{t}^{\alpha}A_{\alpha}^{H}(t)=-\frac{i}{\hbar^{\alpha}}[A_{\alpha}^{H},\mathcal{H}],\label{eq:FHE}
\end{equation}
where we supposed that $U_{\alpha}$ and $\mathcal{H}$ commute.

To build up the deformed Landau-Lifshitz-Gilbert Equation, we use
the eq. (\ref{eq:FHE}), and considering and spin operator $\hat{S}_{\alpha}(t)$,
in such a way that we can write the a deformed Heisenberg equation as

\begin{equation}
D_{t}^{\alpha}\hat{S}_{\alpha}(t)=-\frac{i}{\hbar^{\alpha}}[\hat{S}_{\alpha},\mathcal{H}],
\end{equation} whith

\begin{equation}
\mathcal{H}=-g_{\alpha}\frac{\mu_{B}}{\hbar^{\alpha}}\hat{S}_{\alpha}\circ\vec{H}_{eff}.
\end{equation}

Here, $\vec{H}_{eff}$ is some effective Hamiltonian whose form that we will turn out clear forward.

Now, consider the deformed momentum operator as \cite{nosso AHEP-g-factor,Aspects-nosso,Zitter}

\begin{equation}
\widehat{p}^{\alpha}=-i\left(\hbar\right)^{\alpha}M_{x,\alpha}\frac{\partial^{\alpha}}{\partial x^{\alpha}}.\label{eq:Oper p_alpha}
\end{equation}

Taking this operator, we obtain a deformed algebra, here in terms
of commutation relation for coordinate and momentum 
\begin{equation}
\left[\hat{x}_{i}^{\alpha},\hat{p}_{j}^{\alpha}\right]=\imath\Gamma(\alpha+1)\hbar^{\alpha}M_{\alpha}\delta_{\imath j}I
\end{equation}
and for angular momentum components as 
\begin{equation}
\left[\hat{L}_{i}^{\alpha},\hat{L}_{j}^{\alpha}\right]=\imath\Gamma(\alpha+1)\hbar^{\alpha}M_{\alpha}\hat{L}_{k}^{\alpha}.
\end{equation}

The resulting the $\alpha-$deformed LLG equation can now be written as
 
\begin{equation}
_{0}^{J}D_{t}^{\alpha}\hat{S}_{\alpha}(t)=-\frac{M_{\alpha}\Gamma(\alpha+1)g_{\alpha}\mu_{B}}{\hbar^{\alpha}}\hat{S_{\alpha}}\times\vec{H}_{eff}.\label{eq:Alpha LLG}
\end{equation}

If we take $\hat{m}_{\alpha}\equiv\gamma_{\alpha}\hat{S}_{\alpha}$, $\gamma_{\alpha}\equiv\frac{M_{\alpha}\Gamma(\alpha+1)g_{\alpha}\mu_{B}}{\hbar^{\alpha}}$, we can re-write the equation as the $\alpha-$deformed LLG

\begin{equation}
_{0}^{J}D_{t}^{\alpha}\hat{m}_{\alpha}(t)=-\left|\gamma_{\alpha}\right|\hat{m}_{\alpha}\times\vec{H}_{eff},\label{eq:LLG_m Fract}
\end{equation} with $\vec{H}_{eff}=H_{0}\hat{k}$. 
We have the Solution of eq.(\ref{eq:LLG_m Fract}):

\begin{equation}
m_{\alpha x}=A\cos\theta_{0}E_{2\alpha}(-\omega_{0}^{2}t^{2\alpha})+A\sin\theta_{0}.x.E_{2\alpha,1+\alpha}(-\omega_{0}^{2}t^{2\alpha}).
\end{equation}

In the figure below, the reader may notice the behavior of the magnetization,
considering $\theta_{0}=0$.

\begin{figure}[H]
\includegraphics[clip,width=8cm,height=8cm,keepaspectratio]{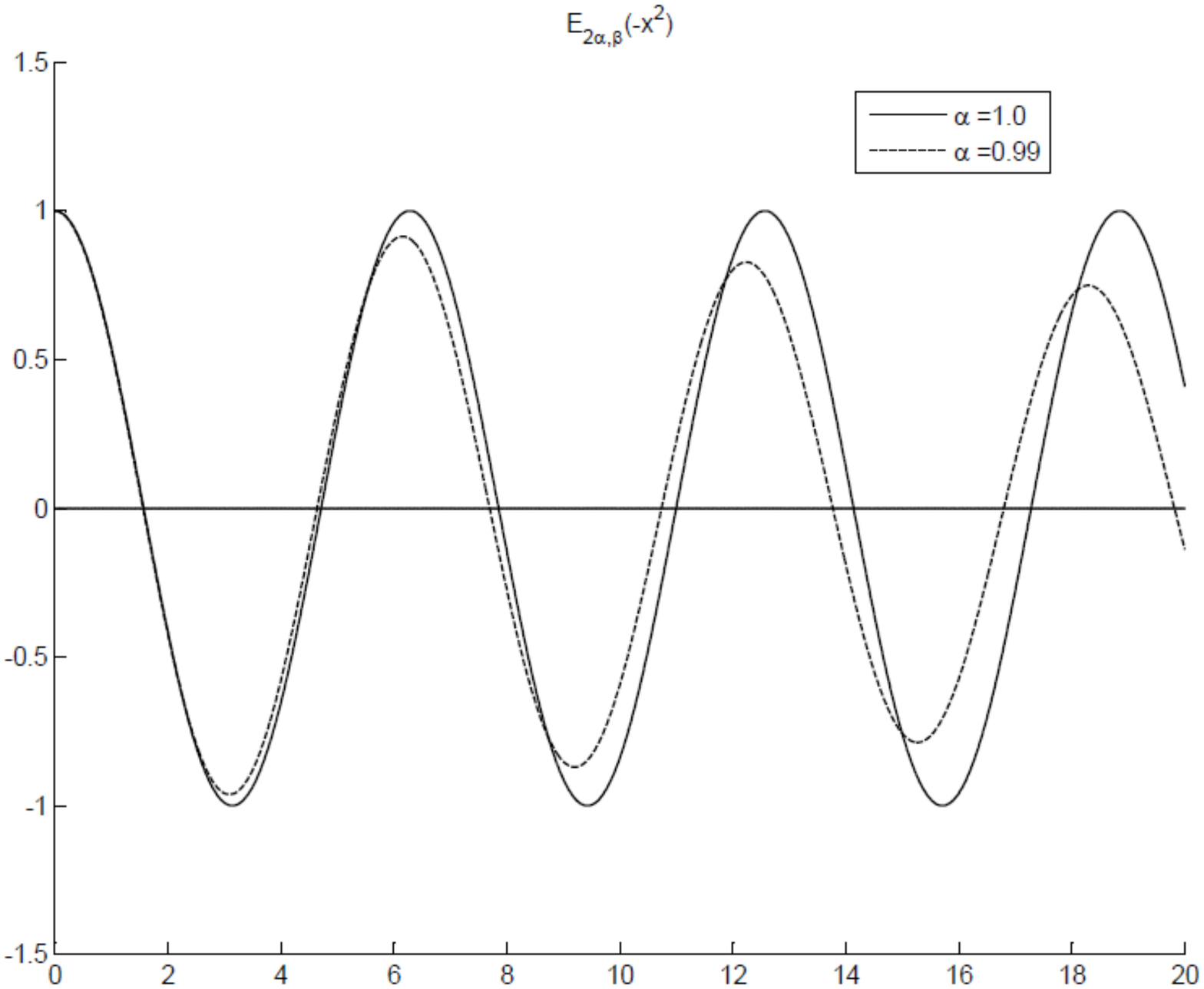}
\includegraphics[clip,width=8cm,height=8cm,keepaspectratio]{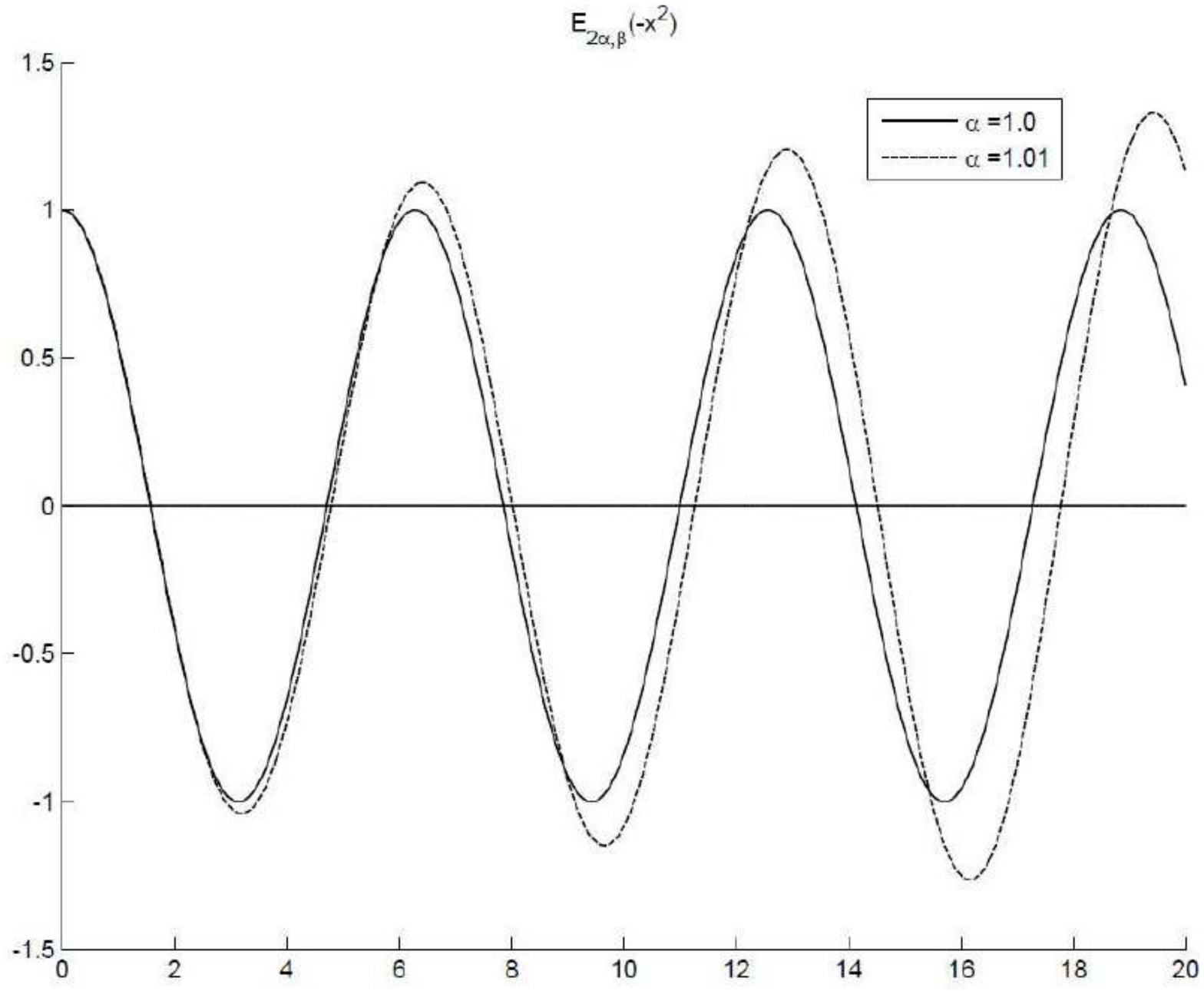}
\\
\caption{ a) Damping of oscillations. In the figure $\beta=1$. b) Increase
of oscillations}.
\end{figure}

For $\alpha=1,$ the solution reduces to $m_{x}=A\cos(\omega_{0}t+\theta_{0})$,
the standard Simple Harmonic Oscillator solution for the precession
of magnetization.

The presence of complex interactions and dissipative effects that
are not explicitly included into the Hamiltonian can be seen with
the use of deformed metric derivatives. Without explicitly adding
up the Gilbert damping term, the damping in the oscillations could
reproduce the damping described by the Gilbert term or could it disclose
some new extra damping effect. Also, depending on the relevant parameter,
the $q-$ entropic parameter or for $\alpha$, the increasing oscillations
can signally that it is sensible to expect fractionality to interfere
on the effects of polarized currents as the Slonczewski term describes.
We point out that there are qualitative similarities in both cases,
as the damping or the increasing of the oscillations, depending on
the relevant control parameters. Despite that, there are also some
interesting differences, as the change in phase for axiomatic derivative
application case.

Here, we cast some comments about an apparent paradox: If we make,
as usually done in the literature for LLG, the scalar product in eq.
(\ref{eq:q-LLG equation for m}) with, $\hat{m}_{q},$ we obtain an
apparent paradox that the modulus of $\hat{m}_{q}$does not change.
On the other hand, if instead of $\hat{m}_{\alpha},$we proceed now
with a scalar product with $\vec{H}_{eff}$ and we obtain thereby
the indications that the angle between $\hat{m}_{\alpha}$and $\vec{H}_{eff}$
does not change. So, how to explain the damping in osculations for
$\hat{m}_{q}?$ This question can be explained by the the following
arguments. Even the usual LLG equation, with the term of Gilbert,
can be rewritten in a form similar to eq. LLG without term of Gilbert.
See eq. (2.7) in the Ref. \cite{The fascinating world of the Landau=00003D002013Lifshitz=00003D002013Gilbert}.
The effective $\vec{H}_{eff}$ field now stores information about
the interactions that cause damping. In our case, when carrying out
the simulations, we have taken $\vec{H}_{eff}$ as a constant effective
field. Here, we can argue that the damping term, eq. (2.8) in Ref.
\cite{The fascinating world of the Landau=00003D002013Lifshitz=00003D002013Gilbert}
being small, this would cause the effective field $\vec{H}_{eff}=\overrightarrow{H}(t)+\overrightarrow{k}(\overrightarrow{S}\times\overrightarrow{H})$
to be approximately $\overrightarrow{H}(t)$. In this way, the scalar
product would make dominate over the term of explicit dissipation.
This could, therefore, explain the possible inconsistency.

\section{Conclusions and Outlook}

In short:

Here, we tackle the problem of LLG equations considering the presence
of complexity and dissipation or other interactions that give rise
to the term proposed by Gilbert or the one by Slonczewski.

With this aim, we have applied scale -$q-$derivative and the axiomatic
metric derivative to build up deformed Heisenberg equations. The evolution
operator naturally emerges with the use of each case of the structural
derivatives. The deformed LLG equations are solved for a simple case,
with both structural or metric derivatives.

Also, in connection with the LLG equation, we can cast some final
considerations for future investigations:

Does fractionality simply reproduce the damping described by the Gilbert
term or could it disclose some new effect extra damping?

Is it sensible to expect fractionality to interfere on the effects
of polarized currents as the Slonczewski term describes?

These two points are relevant in connection with fractionality and
the recent high precision measurements in magnetic systems may open
up a new venue to strengthen the relationship between the fractional
properties of space-time and Condensed Matter systems.

\medskip{}

\thanks{The authors wish to express their gratitude to FAPERJ-Rio de Janeiro
and CNPq-Brazil for the partial financial support.}

\textbf{\bigskip{}
 }


\begin{thebibliography}{99}
\bibitem{Nosso On conection2015}J. Weberszpil, Matheus Jatkoske Lazo
and J.A. Helay\"el-Neto, Physica A 436, (2015) 399\textendash 404.

\bibitem{Variational Deformed}Weberszpil, J.; Helay\"el-Neto, J.A.,
Physica. A (Print), v. 450, (2016) 217-227; arXiv:1511.02835 {[}math-ph{]}.

\bibitem{Tsallis1} C. Tsallis, J. Stat. Phys. {52}, (1988) 479-487.

\bibitem{Tsallis BJP- 20 anos}C. Tsallis, Brazilian Journal of Physics,
39, 2A, (2009) 337-356.

\bibitem{Tsallis2} C. Tsallis, Introduction to Nonextensive Statistical
Mechanics - Approaching a Complex World (Springer, New York, 2009).

\bibitem{Balankin PRE85-Map-2012}Alexander S. Balankin and Benjamin
Espinoza Elizarraraz, Phys. Rev. E 85, (2012) 056314.

\bibitem{Balankin Rapid Comm}A. S. Balankin and B. Espinoza, Phys.
Rev. E 85, (2012) 025302(R).

\bibitem{Balankin-Towards a physics on fractals}Alexander Balankin,
Juan Bory-Reyes and Michael Shapiro, Phys A, in press, (2015) doi:10.1016/j.physa.2015.10.035.

\bibitem{nosso AHEP-g-factor}Weberszpil, J. ; Helay\"el-Neto, J. A.,
Advances in High Energy Physics, (2014), p. 1-12.

\bibitem{NRT-PRL-2011}F. D. Nobre, M. A. Rego-Monteiro, and C. Tsallis,
Phys. Rev. Lett. 106, (2011) 140601.

\bibitem{Aspects-nosso}J. Weberszpil, C.F.L. Godinho, A. Cherman
and J.A. Helay\"el-Neto, In: 7th Conference Mathematical Methods in
Physics - ICMP 2012, 2012, Rio de Janeiro. Proceedings of Science
(PoS). Trieste, Italia: SISSA. Trieste, Italia: Published by Proceedings
of Science (PoS), 2012. p. 1-19.

\bibitem{Zitter}J. Weberszpil and J. A. Helay\"el-Neto, J. Adv. Phys.\textbf{7},
2 (2015) 1440-1447, ISSN 2347-3487.

\bibitem{Axiomatic Metric}J. Weberszpil, J. A. Helay\"el-Neto, arXiv:1605.08097
{[}math-ph{]}

\bibitem{The fascinating world of the Landau=00003D002013Lifshitz=00003D002013Gilbert}M.
Lakshmanan, Phil. Trans. R. Soc. A (2011) 369, 1280\textendash 1300
doi:10.1098/rsta.2010.0319\end{thebibliography}
\end{document}